# MOBAFS: A Multi Objective Bee Algorithm for Feature subset selection in Software Product Lines

Nahid Hajizadeh[1], Peyman Jahanbazi[2], Reza Akbari [3*]

Department of Computer Engineering and Information Technology, Shiraz University of Technology

[1]n.hajizadeh@sutech.ac.ir, [2]p.jahanbazi@sutech.ac.ir, [3]akbari@sutech.ac.ir

___

*Abstract*: Software product line represents software engineering methods, tools and techniques for creating a group of related software systems from a shared set of software assets. Each product is a combination of multiple features. These features are known as software assets. So, the task of production can be mapped to a feature subset selection problem which is an NP-hard combinatorial optimization problem. This issue is much significant when the number of features in a software product line is huge. In this paper, a new method based on Multi Objective Bee Swarm Optimization algorithm (called MOBAFS) is presented. The MOBAFS is a population based optimization algorithm which is inspired by foraging behavior of honey bees. The is used to solve a SBSE problem. This technique is evaluated on five large scale real world software product lines in the range of 1,244 to 6,888 features. The proposed method is compared with the state-of-the-art, SATIBEA. According to results of three solution quality indicators and two diversity metrics, the proposed method, in most cases, surpasses the other algorithm.

___

## 1. INTRODUCTION

Nowadays few software products are produced individually. In fact, most organizations tend to develop families of similar softwares. These products share some common elements which is named software asset. A software asset is a description of a solution or knowledge that application engineers use to develop or modify products in a software product line [1]. By applying reusability concept, shared software assets can be reused, instead of developing from scratch. The process of software asset reusability is what flows in a Software Product Line. Software Product Lines are software systems that share common regulated set of features which include architecture, design, documents, test cases and other assets, also a standard definition of a software feature has been established by IEEE[1]:" *a distinguishing characteristic of a software item (for example, performance, portability, or functionality)* " [2]. Actually, features are employed to develop products. In other word, each product is combination of some features and selecting these features is based on the attributes which are assigned to each feature.

SPL development process is divided into two phases, Domain Engineering and Application Engineering [3]. Domain engineering is the process of defining and realizing the commonality and the variability of the software product line, which refers to the core assets and application engineering process is to develop particular applications by employing the variability of the software product line [3]. In other words, application engineering is product generation using the core asset achieved by the domain engineering, which is mentioned in this article.

A Feature Model or Feature Diagram is a compact display of all products in terms of features in a software product line. Indeed, a feature model shows the principle features of a product's family in the domain and relationships between them [4]. On the other hand, a feature model is a compressed demonstration of all the products of SPL in respect of "features". A feature model has a tree structure which defines the relationships between features in a hierarchical pattern.

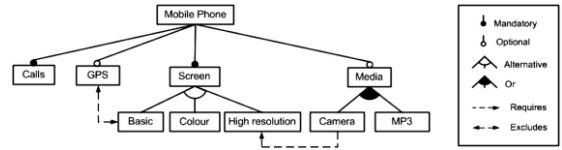

Figure 1: A sample feature model for mobile phone product line [5]

In Figure 1, a feature model based on an instance software product line for mobile phone is illustrated. In a feature model there are two types of relationship between features which indicate:

- A parent feature and its children features relationships
- Cross-tree constraints

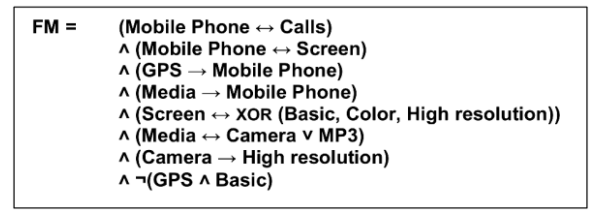

Figure 2: Feature model of mobile phone SPL in a Boolean expression format [6]

---
[1] The Institute of Electrical and Electronics Engineers

A feature model can be expressed in Boolean expression format, as depicted in Figure 2.

Software products are created by merging the selected features from a feature model and considering all the constraints. Here the issue is to select and extract a set of features from a feature model and the goal is selecting the features set in a way that not only cover restrictions but also be optimized in terms of the objective functions. Feature models consist of hundreds and thousands features on an industrial scale so it is practically impossible to use exact algorithms to derive products in a reasonable time. In other words, how to combining a set of features to make a product, in an optimal manner, is an NP-hard problem, as White et al.[7] indicated, and is known as a feature subset selection problem in SPLs. Therefore, the need for a heuristic algorithm for solving such problems is felt. Hence, in this paper a new method based on Multi Objective Bee Swarm Optimization (MOBSO) algorithm to get an optimal solution of the problem in an acceptable time is presented.

Dealing with combinatorial optimization problems, in most cases, is time consuming so this kind of problems has been an alive area of research for many decades. According to this point that optimization problems, in reality, become more complicated, the need for more appropriate optimization algorithms are sensed regularly. In such problems, the objective is finding the minimum or maximum of the function which is related to these problems.

Optimizing more than one objective simultaneously (according to this point that some objects might have conflict) is known as Multi-Objective Optimization [8]. The outputs of Multi-Objective Optimization are solutions which are optimal or near optimal. Pareto Front is a set of non-dominated solutions, being chosen as optimal, if no objective can be improved without sacrificing at least one other objective. On the other hand, a solution x* is referred to as dominated by another solution x if, and only if, x is equally good or better than x* with respect to all objectives and for at least one objective, x is strictly better than x*[9].

With the emergence of swarm intelligence [10], a new solution to optimization problems has been created, which is called swarm-based optimization algorithms. These algorithms are inspired by natural behavior of social animals when searching for food at their own nearest location. The most famous of these algorithms can be mentioned in the following: Particle Swarm Optimization (PSO) [11], Ant Colony Optimization (ACO) [12] and Artificial Bee Colony Algorithm (ABC) [13]. Many algorithms which have been developed and employed to different engineering area are gotten from intelligent behaviors of honey bees [14] [15] [16] [17] [18] [19] [20]. But a few algorithms in this field are applied for numerical optimization problems, as an example can be named Artificial Bee Colony (ABC). In this paper one of the effective algorithm based on foraging behaviors of honey bees is applied, Multi Objective Bees Swarm Optimization (MOBSO), which is proposed by Akbari et al. [21].

MOBSO uses distinct types of bees to optimize numerical functions. Each type of bees uses a different moving pattern. Today, MOBSO has been applied in different scientific areas including Energy management [22], Image enhancement[23] and Clustering [24].

In this article, an algorithm based on MOBSO which we call it MOBAFS is introduced. The results show that the proposed algorithm beats previous most successful algorithm (i.e. SATIBEA) in the majority of performance metrics.

The rest of the paper is organized as follows: in Section 2, related works are introduced. Section 3 describes the MOBAFS algorithm. Section 4 and 5 include experimental setup and experimental results; and finally Section 6 presents some conclusion and future works.

## 2. RELATED WORK

In the field of feature subset selection problem, the methods could be divided to two groups:

- Single Objective
- Multi Objective

### 2.1. Single Objective Approaches

In 2005, Benavides et al. [25] introduced a method which mapped feature selection problem in SPLs to a Constraint Satisfaction Problem (CSP). They experimented their method on four problems (two academicals and two real product lines). Their implementation showed an exponential behavior when the number of features in the feature models were increased. White et al. [7], in 2008, proposed a polynomial time approximation technique which called Filtered Cartesian Flattening (FCE) to achieve approximately optimal solution by transforming the feature selection problem with resource constraints into equivalent Multidimensional Multiple choice Knapsack Problem (MMKP) and applying MMKP approximation algorithms. Their method showed 93% optimality on feature models with 5,000 features. Another researcher which used CSP is White et al. [26] , who proposed a formal model of multi-step SPL, in 2009, and mapped this model to CSPs and called MUlti step Software Configuration probLEm (MUSCLE). An artificial intelligence approach based on genetic algorithms called GAFES which was applied by Guo et al. [27] in 2011 as a search based technique to solve the optimized feature selection in SPLs. GAFES can produce solutions with 86-97% optimality. In 2011, Soltani et al. [28], introduced a framework which uses an artificial intelligence planning technique to select features that satisfy stakeholder's business concern and resource constraints.

## 2.2. Multi Objective Approaches

Sayyad et al. [29], used Multi-objective Evolutionary Optimization Algorithm (MEOA) to solve feature selection problem in SPLs. MEOA can achieve an acceptable configuration for a large feature model (290 features) in as little as 8 minutes when other algorithms had found one acceptable configuration after 3 hours. Later, in November 2013, Sayyad et al. [6] presented simple heuristics to solve the software product lines configuration problem, in the case of multi-objective, which led Indicator-Based Evolutionary Algorithm (IBEA) to find optimum configurations of large scale models. They applied "seed" technique in order to generate initial population randomly and could find 30 sound solutions for configuring a set of 6000 features within 30 minutes.

Olaechea et al. [30] compared an exact and an approximate algorithm in terms of accuracy, time consumption, scalability and parameter setting requirements for solving the software product lines configuration problem on five case studies. Based on their experimental results, they claimed that using exact techniques for small software product lines multi-objective optimization problems are possible and also approximate methods can be applied for large scale problems, however, need considerable effort to find the best parameter setting for a satisfactory approximation.

Tan et al. [31], presented a new approach by introducing a feedback-directed mechanism into various EAs. Their method is based on analyzing violated constraints, and uses the analyzed results as a feedback to guide the process of crossover and mutation operators. However, for Linux repository, which contains 6888 features, their method couldn't find any correct solution.

In May 2015, Henard et al. [32] proposed SATIBEA, a search-based feature subset selection algorithm for SPLs. SATIBEA is mixed of Indicator-Based Evolutionary Algorithm (IBEA) [33] and satisfiability (SAT) solving technique. They considered 5 objects and evaluated SATIBEA on 5 huge real-world SPLs. Their significant results encouraged us to evaluate our approach with SATIBEA.

Recently, Hierons et al.[34] and Xue et al.[35] proposed new approaches based on IBEA. The point is that none of these articles have evaluated their works with considering SATIBEA. In spite of the presence of SATIBEA and proof of to be more powerful than IBEA, both articles compared their methods with IBEA. Another point is that Hierons et al. applied a real repository with maximum 290 features and a randomly generated feature model with 10,000 features. Both articles didn't evaluate their results by common metrics specialized for multi-objective optimization algorithms. Due to significant results of SATIBEA, in comparison with other algorithms such as IBEA, we would prefer to compare our method with that.

As the case of "test case selection" can be seen as a special situation of multi objective product/configuration selection in feature models, so, some articles in the field of software testing are deserved to be introduced.

Parejo et al. [36] in 2016, applied the NSGA-II evolutionary algorithm to solve the multi-objective test case prioritization problem. They proposed seven objective functions based on functional and non-functional data and found that multi-objective prioritization results in fault detection is faster than mono-objective prioritization.

Galindo et al. [37] presented a variability-based testing approach to derive video sequence variants to test different input combinations when developing video processing software. Combinational and multi-objective optimization testing techniques over feature models have been presented to generate a minimized number of configurations which is combinations of features to synthesize variants of video sequences.

SPL pairwise testing is what Lopez-Herrejon et al. [38] have taken into consideration. They applied classical multi-objective evolutionary algorithms such as NSGA-II, MOCell, SPEA2 and PAES to select a set of products to test which maximize the coverage and minimize the test suite size.

Pereira et al. [39] introduced a new method to configure a product that considers both qualitative and quantitative feature properties. They modeled the product configuration task as a combinatorial optimization problem. Their research was the first work in the literature that considers feature properties in both leaf and non-leaf features.

In 2018, Abbas et al. [40] proposed a multi-objective optimum algorithm that consists of three independent paths (first, second, and third). They applied heuristics on these paths and found that the first path is infeasible due to space and execution time complexity and the second path reduces the space complexity. They calculated the outcomes of all three paths and proved the significant improvement of optimum solution without constraint violation occurrence.

Xue and Li [41] exposed the mathematical nature of optimal feature selection problem in software product line and tried to implement three mathematical programming approaches to solve this problem at different scales. The empirical results shown that their proposed method can find significantly more non-dominated solutions in similar or less execution time, in comparison with IBEA.

Yu et al. [42] proposed six hybrid algorithms which combine the SAT solving with different MOEAs. Their case study was based on five large-scale, rich-constrained and real-world SPLs. Empirical results demonstrated that SATMOCell algorithm obtained a competitive optimization performance in comparison with the state-of-the-art that outperformed the SATIBEA in terms of quality

Hypervolume metric for 2 out of 5 SPLs within the same time budget.

Shi et al. [43] introduced a parallel portfolio algorithm, IBEAPORT, which designs three algorithm variants by incorporating constraint solving into the indicator-based evolutionary algorithm in different ways and performs these variants by utilizing parallelization techniques. Their approach utilized the exploration capabilities of different algorithms and improves optimality as far as possible within a limited time budget.

In 2019, Khan et al. [44] proposed a new feature selection method that supports multiple multi-level user defined objectives. A new feature quantification method using twenty operators, capable of treating text-based and numeric values and three selection algorithms called Falcon, Jaguar, and Snail are proposed. Falcon and Jaguar are based on greedy algorithm while Snail is a variation of exhaustive search algorithm. With an increase in 4% execution time, Jaguar performed 6% and 8% better than Falcon in terms of added value and the number of features selected.

Wägemann et al. [45] introduced ADOOPLA, a tool-supported approach for the optimization of product line system architectures. In contrast to existing approaches where product-level approaches only support product-level criteria and product-line oriented approaches only support product-line-wide criteria, their approach integrates criteria from both levels in the optimization of product line architectures. Also, the approach could handle multiple objectives at once, supporting the architect in exploring the multi-dimensional Pareto-front of a given problem.

Xue et al. [46] introduced a new aggregation-based dominance (ADO) for Pareto-based evolutionary algorithms to direct the search for high-quality solutions. Their approach was tested on two widely used Pareto-based evolutionary algorithms: NSGA-II and SPEA2+SDE and validated on nine different SPLs with up to 10,000 features and two real-world SPLs with up to 7 objectives. Their experiments have shown the effectiveness and efficiency of both ADO-based NSGA-II and SPEA2+SDE.

Xiang et al. [47] addressed the open research questions, how different solvers affect the performance of a search algorithm, by performing a series of empirical studies on 21 features models, most of which are reverse-engineered from industrial software product lines. They examined four conflict-driven clause learning solvers, two stochastic local search solvers, and two different ways to randomize solutions. Experimental results suggested that the performance could be indeed affected by different SAT solvers, and by the ways to randomize solutions in the solvers. Their research served as a practical guideline for choosing and tuning SAT solvers for the many-objective optimal software product selection problem.

In 2020, Saber et al. [48] we present MILPIBEA, a novel hybrid algorithm which combines the scalability of a genetic algorithm (IBEA) with the accuracy of a mixed-integer linear programming solver (IBM ILOG CPLEX). We also study the behaviour of our solution (MILPIBEA) in contrast with SATIBEA, a state-of-the-art algorithm in static software product lines.

Lu et al. [49] introduced a pattern-based, interactive configuration derivation methodology, called Pi-CD, to maximize opportunities of automatically deriving correct configurations of CPSs by benefiting from pre-defined constraints and configuration data of previous configuration steps. Pi-CD requires architectures of CPS product lines modeled with Unified Modeling Language extended with four types of variabilities, along with constraints specified in Object Constraint Language (OCL). Pi-CD is equipped with 324 configuration derivation patterns that they defined by systematically analyzing the OCL constructs and semantics.

Hierons et al. [50] proposed a new technique, the grid-based evolution strategy (GrES), which considers several objective functions that assess a selection or prioritization and aims to optimize on all of these. The problem was thus a many-objective optimization problem. They used a new approach, in which all of the objective functions are considered but one (pairwise coverage) was seen as the most important. They also derived a novel evolution strategy based on domain knowledge.

3. THE PROPOSED ALGORITHM

The description of the proposed algorithm for feature subset selection in software product line is given in this section. A set of meta-heuristic algorithms are utilized to solve problems with exponential time complexity which are inspired from honey bees algorithms. The proposed algorithm is designed based on the lesson learned from multi-objective variants of bee algorithms such MOBSO was basically proposed in [21]. This algorithm is a population based optimization technique which is inspired by foraging behavior of honey bees.

This algorithm involves three types of bees; experienced forager, onlooker and scout bees which fly in an $D$-dimensional search space $S \subset R^D$ to find the near optimal solution. Each type of bees has a specific moving pattern which is used by the bees to adapt their flying direction. Experienced foragers use an adaptive windowing mechanism to select their leaders and regulate their next positions. In addition, for cutting the most crowded members of the archive, the adaptive windowing mechanism is applied too. In MOBAFS Scouts and adapting windowing mechanism maintain diversity over the Pareto front. The structure of the algorithm is shown in Figure 3.

The MOBAFS input parameters are the bee population size, maximum iteration and the maximum number of non-dominated bees. *Initialization* is the first phase of the algorithm where number of bees is randomly generated and also non-dominated bees are determined. The second phase, *Update*, is a loop with max_iter iteration. In this phase, bees move according to their pattern in the search space. In each iteration, type of bee is specified. Then the experienced forager bees, onlooker bees, and scout bees move in the search space with specific patterns of movement. Finally, if the number of solutions in archive exceeds the arch size, some of elements will be deleted. Further details about each step of the algorithm are given as follows.

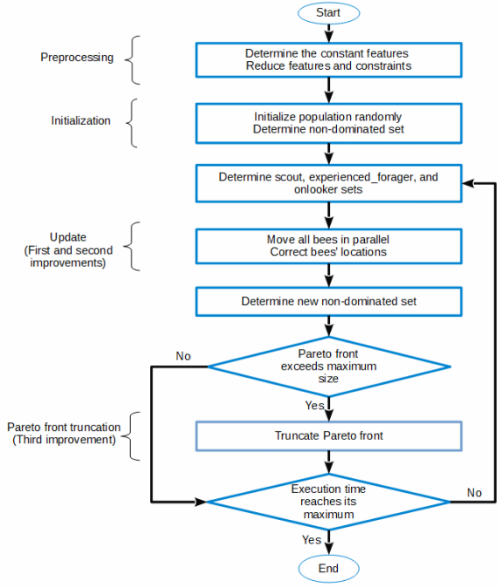

Figure 3-MOBAFS algorithm

### 3.1. Initilization

Figure 4 illustrates the initialization part of MOBAFS algorithm.

bees = **random**(D);
arch = **add_non_dominated**(null, bees);

Figure 4: MOBAFS algorithm-Initialization step

In this step, two sets *bees* and *arch* are initialized which indicate bees set and non-dominated bees set respectively. *Random()* function creates a random point in the *D*-dimensional space. In addition, the *add_non_dominated()* function receives two sets as the input (*arch* and *bees'* sets) and extracts non-dominated set from inputs.

### 3.2. Update

This step is repeated by *max_iter*. At first, the type of bees is determined. Then each bee moves in the search space according to its type and finally *arch* set is updated. Figure 5 illustrates these steps.

scout = **get_scout**(bees, ps);
experienced_forager = **select_non_dominated**(arch, bees) - scout;
onlooker = bees - Experienced _forager – scout;

Figure 5: MOBAFS algorithm-Update step

As shown in Figure 5, Scout bees are determined by calling *get_scout()* function. The *get_scout()* function chooses *ps* percent of bees randomly. In the next step, *select_non_dominated()* function determines experienced forager bees. This function intersects *arch* and *bees'* set to calculate experienced forager bees. The result subtracted from Scout set. Lastly, onlooker bees are determined, the bees who are neither experienced forager nor scout.

*Forager movement:* Figure 6 shows the pattern of experienced forager bees' movement.

**For all** bee **in** experienced_forager
**Begin**
  leader = **select_leader**(arch);
  bee = bee + $w_l$ + $r_l$ ∗ (leader − bee)
**End**

Figure 6: Experienced forager bees movement

In the first step, a forager bee determines a leader bee. The leader bee is randomly selected from *arch* set. The bees in the less crowded location in search space has more probability to be selected as a leader. The details of this process is explained in [21]. In the next step, new position of the bee is calculated, according to two parameters $w_L$ and $r_L$ which controls the importance of the information provided by the leader and a random variable to uniform distribution in range [0.1], respectively.

*Onlooker movement:* The way onlooker bees move is shown in Figure 7.

**for all** bee in onlooker
**Begin**
  elite = **select_elite**(experimced_forager);
  x = x + $w_e$ ∗ $r_e$ ∗ (elite − bee)
**End**

Figure 7: Onlooker bees movement

The movement pattern of onlooker bees is very similar to forager bees' movement. The difference is that each bee randomly selects an elite bee from forager bees. The new position of an onlooker bee is determined based on two parameters $w_e$ and $r_e$ which controls the importance of the information provided by the elite and a random variable to uniform distribution in range [0.1], respectively.

*Scout movement:* The pattern of scout bees' movement is presented in Figure 8.

**for all** bee in scout
**Begin**
  $b_1$ = **select_random**(arch);
  $b_2$ = **select_random**(arch);
  bee = **move_randomly**($b_1$, $b_2$);
**End**

Figure 8: Scout bees movement

The Scout bees' movement begins with a random selection of two bees from the *arch* set as lower and upper bounds to represent the search space and then a Scout bee move in this space.

The last part of *Update* step is updating the archive. At first, from the *bees* and *arch* sets, non dominated solutions are selected and stored in *arch*. If the number of *arch*'s elements is over than *arch_size*, *truncate_archive()* function will eliminate extra element. The bees in the most crowded locations have more probability to be removed.

---
arch = **add_non_dominated**(arch, bees);
**if** n(arch) > arch_size **then**
   **truncate_archive**(arch);
**End**

---

Figure 9: Truncate Pareto front

In order to improve the efficiency of the MOBAFS algorithm and create more optimized solutions, three developments on MOBAFS have been done which will be described in following.

The second development is related to generate solutions with less constraints violation. For this purpose, some changes have been done in movement functions. In all three movement functions (experienced forager, onlooker and scout) according to constant features, the positions of bees are repaired. After that, the SAT (Satisfiability) solver checks whether any constraint has been violated due to the new positions. In case of constraints violation, some features are selected randomly and their statuses are changed to indeterminate, then the SAT solver tries to assign proper values to indeterminate features in order to eliminate constraints violation. This process continues until the SAT solver notifies that no constraints are violated. In this way the appropriate value for features and subsequently the new position of bees will be determined. Figure 10 illustrates how the SAT solver achieves a solution without any constraint violation. In this work to generate valid configurations, Sat4j [51], a SAT solver which is one of the most popular one, is applied.

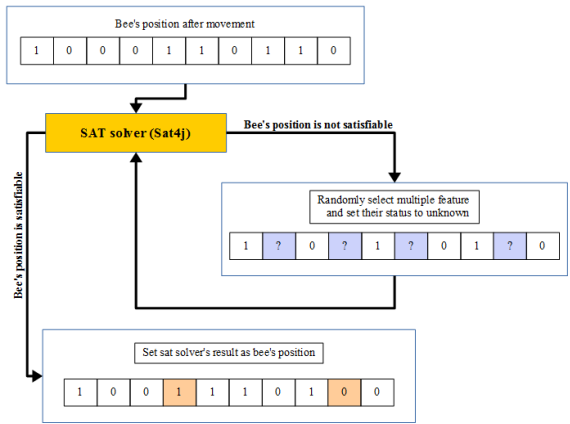

Figure 10: An example of SAT solver role in repositioning bees

The third development is related to the *truncate_archive()* function which its modified version is shown in Figure 11. This function is called, same as previous version, at each iteration of MOBAFS. In modified version of this function, two points are considered:

1) Defining a process that can compare two non-dominated solutions to avoid accidental removal of high quality solutions.
2) Trying to keep solutions that are located in non-crowded location of state space.

This function adds two attributes, GROUP and RANK, to each node. The GROUP attribute helps to determine crowded positions and RANK attribute orders non-dominated solutions based on the quality of those solutions. Finally, in each GROUP, in the number of RANK of the solution, adjacent solutions stay in *arch* set. Figure shows this function in more detail.

To make the algorithm's steps more clear, a flowchart which is shown in Figure 3 is presented.

*3.3. Speedup*

In any algorithm, one of the most important parameters is evaluating the performance of it in terms of time processing and memory consumption. In this section, our MOBAFS algorithm with regard to time complexity is investigated.

If the total execution (in the case of serial implementation), initialization, loop iterations and Pareto Front truncation time be $T_{total}, T_{init}, T_{loop}$ and $T_{trunc}$, respectively, then the total processing time is calculated according to Eq.1:

$$T_{total\,(serial)} = T_{init} + T_{loop} + T_{trunc} \quad (1)$$

All the times mentioned in Eq.1 is measured through Eq.2, Eq.3 and Eq.4 where $I$ is the number of repetition of iterations loop, $N$ is the population size, $F$ is the number of features and $C$ is the number of constraints (In these formulas, for simplicity, the constant factors are not mentioned):

$$T_{init} = N(F + C) + N^2 \quad (2)$$

$$T_{loop} = I * N(F + C + N) \quad (3)$$

$$T_{trunc} = N^2 \quad (4)$$

According to Eq.2 to Eq.4, the total time complexity of MOBAFS algorithm that is shown in Eq.1 can be rewritten as Eq.5:

$$T_{total(serial)} = I * N(F + C + N) \quad (5)$$

With the aim of parallelism, the total time complexity is reduced to Eq.6 where $P$ is the number of processors:

$$T_{total(parallel)} = I * N \left(\frac{F+C}{P} + N\right) \quad (6)$$

Consequently, Eq.7 shows the speedup, based on the Eq.5 and Eq.6:

$$Speedup = \frac{T_{total(serial)}}{T_{total(parallel)}} = \frac{I*N(F+C+N)}{I*N\left(\frac{F+C}{P}+N\right)} \quad (7)$$

**Algorithm: Truncate Pareto front**

**Input**:
    archive: list of non-dominated solutions
    sp: number of section per each fitness function
**Output**:
    newArchive: Minified list of non-dominated solutions

**Function** truncateArchive (archive, sp)
**Begin**
    newArchive = {};
    **For** all x **in** archive
    **Begin**

$$a = \left\lfloor \frac{sp * (x \cdot correctness - \min(correctness))}{\max(correctness) - \min(correctness)} \right\rfloor$$

$$b = \left\lfloor \frac{sp * (x \cdot richness - \min(richness))}{\max(richness) - \min(richness)} \right\rfloor$$

$$c = \left\lfloor \frac{sp * (x \cdot usedBefore - \min(usedBefore))}{\max(usedBefore) - \min(usedBefore)} \right\rfloor$$

$$d = \left\lfloor \frac{sp * (x \cdot knowsDefects - \min(knowsDefects))}{\max(knowsDefects) - \min(knowsDefects)} \right\rfloor$$

$$e = \left\lfloor \frac{sp * (x \cdot cost - \min(cost))}{\max(cost) - \min(cost)} \right\rfloor$$

$$x \cdot group = a + 255 * \big(b + 255 * \big(c + 255 * (d + 255 * e)\big)\big)$$

$$x \cdot rank = (sp + 1 - a)(sp + 1 - b)(sp + 1 - c)(sp + 1 - d)(sp + 1 - e)$$

    **End**
    $T = \{x \in 2^{archive} | \forall m \& n \in x \quad m \cdot group = n \cdot group \& (\forall t \in T \; x \not\subset t \; or \; t = x )\}$
    **For** all t **in** T
    **Begin**
        R = rank(t)
        Randomly select R member of t set and add to newArchive
    **End**
    **Return** newArchive
**End**

Figure 11- Truncate Pareto front

The variables in Figure 11 are defined as:

$x \cdot correctness$: The Correctness fitness function of Solution x
$x \cdot richness$: The richness fitness function of Solution x
$x \cdot usedBefore$ : The used before fitness function of Solution x
$x \cdot knowsDefects$: The known defect fitness function of Solution x
$x \cdot cost$: The cost fitness function of Solution x
$\max(correctness) = t \cdot correctness \mid t \in archive$ and $\forall x \in archive \; t \cdot correctness \geq x \cdot correctness$
$\max(richness) = t \cdot richness \mid t \in archive$ and $\forall x \in archive \; t \cdot richness \geq x \cdot richness$
$\max(usedBefore) = t \cdot usedBefore \mid t \in archive$ and $\forall x \in archive \; t \cdot usedBefore \geq x \cdot usedBefore$
$\max(knowsDefects) = t \cdot knowsDefects \mid t \in archive$ and $\forall x \in archive \; t \cdot knowsDefects$
        $\geq x \cdot knowsDefects$
$\max(cost) = t \cdot cost \mid t \in archive$ and $\forall x \in archive \; t \cdot cost \geq x \cdot cost$
$\min(correctness) = t \cdot correctness \mid t \in archive$ and $\forall x \in archive \; t \cdot correctness \leq x \cdot correctness$
$\min(richness) = t \cdot richness \mid t \in archive$ and $\forall x \in archive \; t \cdot richness \leq x \cdot richness$
$\min(usedBefore) = t \cdot usedBefore \mid t \in archive$ and $\forall x \in archive \; t \cdot usedBefore \leq x \cdot usedBefore$
$\min(knowsDefects) = t \cdot knowsDefects \mid t \in archive$ and $\forall x \in archive \; t \cdot knowsDefects$
        $\leq x \cdot knowsDefects$
$\min(cost) = t \cdot cost \mid t \in archive$ and $\forall x \in archive \; t \cdot cost \leq x \cdot cost$

## 4. EXPERIMENTAL SETUP

In this section, the experiments that have been done to evaluate the performance of the MOBAFS algorithm on a number of data sets are represented. The performance of the MOBAFS algorithm is evaluated in comparison with SATIBEA [32]. The comparisons are performed based on the famous metrics which have been broadly used to compare optimization algorithms [13] [52] [53] [54]. They are divided in two groups; Quality and Diversity metrics. These metrics and their definitions has been listed in Table 1.

TABLE 1- THE METRICS AND THEIR DEFINITIONS

| | Metric | Definition |
|---|---|---|
| **Quality** | Hypervolume (HV) [55] | It evaluates how well a Pareto front fulfills the optimization objectives. |
| | Epsilon ($\varepsilon$) [56] | It measures the shortest distance which is required to transform every solution in a Pareto front to dominate the reference front. |
| | Inverted Generational Distance (IGD) [57] | It is the average distance from the solutions owned by the reference front to the closest solution in a Pareto front. |
| **Diversity** | Pareto Front Size (PFS) | It is the number of solutions in a Pareto front. |
| | Spread (S) [58] | It determines the amount of spread in Pareto front's solutions. |

Each algorithm was run 30 times per feature model and each run lasted 30 minutes for execution time.

### 4.1. Solution Modeling

The first step for applying MOBAFS algorithm to any problem is to represent a solution as a point in a *D*-dimensional search space. Suppose a problem with $F$ features is presented. To represent one solution, state of $F$ features must be determined. Each feature can have one of two situations; selected (1) or not (0). In this case, one solution will be shown with $F$ bits. For transforming the $F$ bits to integers (or real), each 32 features can be displayed with an integer number. Consequently displaying $F$ features need $\left\lceil\frac{F}{32}\right\rceil$ integer. In other word, domain of problem will be $\mathbb{N}^{\left\lceil\frac{F}{32}\right\rceil}$.

### 4.2. Data Set

The feature models, which are used in this article, are taken from the LVAT (Linux Variability Analysis Tools) repository [2]. The 5 feature models and their characteristics (the version, the number of features and constraints) are listed in Table 2.

TABLE 2- FEATURE MODELS [32]

| Feature Model | Version | Features (mandatory) | Constraints |
|---|---|---|---|
| Linux | 2.6.28.6 | 6,888 (58) | 343,944 |
| uClinux | 20100825 | 1,850 (7) | 2,468 |
| Fiasco | 2011081207 | 1,638 (49) | 5,228 |
| FreeBSD | 8.0.0 | 1,396 (3) | 62,183 |
| eCos | 3.0 | 1,244 (0) | 3,146 |

### 4.3. Preprocessing

As already indicated in section 1, feature selection is based on the attributes of each feature. Due to augmentation that has been done by Sayyad et al.[59],[29] and Henard et al. [32] three attributes is added to each feature: cost, used before and defects.

[2] http://code.google.com/p/linux-variability-analysis-tools

The values of these 3 attributes are set randomly. *Cost* is a real number in range of 5.0 to 15.0, *Used before* is a Boolean variable and *Defects* is an integer in range of 0 to 10. These 3 attributes have a dependency among them: if (not used before) then defects=0.

Another major preprocessing performed includes reduce in dimensions of the problem domain. In previous section, it is mentioned that a problem with $F$ features can be indicated by a point in space with size $\frac{F}{32}$. The smaller value for $\frac{F}{32}$ there is the smaller search space. Given that the value of F depends on the type of problem and its value cannot be changed, so only the value of $\frac{F}{32}$ has to be changed. In order to reduce the search space, with considering the constraints of any problem, the status of some features can be determined before the execution of the algorithm. In fact, the status of those features can be identified before running the algorithm that:

1) Those features are mandatory
2) Due to the status of mandatory features and constraints set, their values can be determined. These features are called constant features, although in [6] called "fixed features".

---

**Algorithm: Determining constant features**

**Input**:
    Constraints: A CNF.

**Output**:
    zero and one: Two arrays of constant features that zero is array of variables with false value and one is an array of variables with true value.

---

```
zero = {};
one = {};
defined = {};
Changed = true;
while (changed = true)
Begin
 changed = false;
 for (c in constraints)
 Begin
  for (x in c)
  Begin
   others = c - {x};
   if  x ∉ defined   and  (∀ t ∈ others |t ∈ zero ∨
 -t ∈ one) then
    Begin
     if x > 0 then
      one = one ∪ {x}
      defined = defined ∪ {x}
     else
      zero = zero ∪ {-x}
      defined = defined ∪ {-x}
    End
    changed = true;
  End
 End
End
```

Figure 12- Determining constant features

To clarify these two modes, two examples are presented.

The first mode: If $p$ is a mandatory feature then before running the algorithm its value can be considered (1), i.e. *Enabled*.

The second mode: if among constraints there is a constraint in form $\sim p \vee q$, according to the value of $p$ (here $p$ is a mandatory feature so its value is considered 1 ), in order to satisfy the constraint $\sim p \vee q$, the value of $q$ has to be (1).

Figure 12 shows the function of constant and mandatory feature selection. In fact, by executing this function constant features are specified before executing MOBAFS. By executing the proposed function, the dimensions of the search space will be reduced from $\frac{F}{32}$ to the formula (8), where M is the number of mandatory features and t is the number of constant features.

$$\frac{F - M - t}{32} \quad (8)$$

Table 3 consists of feature models, the number of features, constant features, constraints and declined constraints.

TABLE 3- FEATURE MODELS WITH DECLINED CONSTRAINTS

| Feature Model | Features | Constant Features | Constraints | Declined Constraints |
|---|---|---|---|---|
| Linux | 6,888 | 154 | 343,944 | 192 |
| uClinux | 1,850 | 1244 | 2,468 | 1256 |
| Fiasco | 1,638 | 1013 | 5,228 | 1059 |
| FreeBSD | 1,396 | 4 | 62,183 | 6 |
| eCos | 1,244 | 23 | 3,146 | 58 |

As it is evident, by using this technique, on average, 26.69 percentage of feature status and 14.61 percentage of constraints are determined before searching the search space. Determining the features status before searching the search space, in addition to reducing the search space has two more advantages:

1) The unfeasible solutions were not searched
2) Decline in the problem's constraints

The first advantage seems obvious because by determining the value of each feature before execution, that features can't be assigned to any other value during execution. For second advantage, it can be said that if the value of all the features of a proposition be determined in a way that the proposition has right value, and also the value of features does not change during execution then that proposition will be right for ever and it's not necessary to re-evaluate it.

### 4.4. Intended Optimization Objectives

According to this point that Henard et al. [32] applied 5 objects for this issue and this article tends to evaluate its performance in comparison with the mentioned paper, so in this work 5 objects are considered too which are listed in Table 4 :

TABLE 4- THE OBJECTIVES AND THEIR OPTIMAL SITUATIONS

| Objective | Optimal situation |
|---|---|
| Correctness | minimizing the violated constraints |
| Richness of features | minimizing the number of deselected features |
| Used before | minimizing the features that were not used before |
| Defects | minimizing the number of defects |
| Cost | minimizing the cost |

### 4.5. MOBAFS Configuration

In this work, the MOBAFS algorithm is adjusted with 3000 populations and 30 independent runs has been done. The reason for considering 3000 populations, contrary to what is common and less than this amount, is that by creating more initial population, the probability of finding more non-dominated solutions increases. The number of experienced forager and onlooker bees are 98 percent of the population which is divided between them at each iteration of the algorithm dynamically and consequently scout bees are 2 percent of the population. The weighting coefficients $w_l$ (the parameter which controls the importance of the knowledge provided by the leader bee; a randomly selected bee from bees' archive) and $w_e$ (the parameter which controls the importance of the knowledge provided by the Experienced forager bees) are adapted to 2.5 and 2.12 respectively. All experiments were performed on Ubuntu 16.04 LTS 64bit with Intel Core i7 4790K CPU 4GHz and 16GB RAM.

### 5. EXPERIMENTAL RESULTS

The MOBAFS and SATIBEA were run on the five feature models, listed in Table 3, to evaluate their performance. In Table 5, the results based on 5 metrics, indicated in Table 4, and 30 independent runs for each algorithm in 30 minutes are presented. The values of measured metrics are the average values in 30 runs.

As it is mentioned in Table 1, the Hypervolume (HV) metric shows the volume of the area which is dominated by a solution set. Therefore, the Pareto front of an algorithm with higher HV is selected more precisely.

Figure 10 presents Hypervolume of MOBAFS and SATIBEA. It is obvious that these two algorithms in two feature models, which are the densest in terms of the number of constraints, i.e., Linus and FreeBSD, have approximately same values, though the HV values of these two algorithms for other feature models have miner difference. So it can be said that the power of these two algorithms in Hypervolume phase are roughly equal.

TABLE 5-EXPERIMENTAL RESULTS IN COMPARISON WITH MOBAFS AND SATIBEA.

| | | | MOBAFS | SATIBEA |
|---|---|---|---|---|
| Linux | Quality | HV | 2.35E-01 | 2.38E-01 |
| | | ε | 1.51E+00 | 9.90E-01 |
| | | IGD | 1.66E-03 | 5.52E-03 |
| | Diversity | PFS | 2.61E+03 | 2.91E+02 |
| | | S | 6.43E-01 | 1.25E+00 |
| uClinux | Quality | HV | 2.41E-01 | 2.76E-01 |
| | | ε | 2.01E-01 | 1.31E+00 |
| | | IGD | 1.49E-03 | 1.35E-02 |
| | Diversity | PFS | 7.96E+02 | 3.00E+02 |
| | | S | 6.96E-01 | 1.37E+00 |
| Fiasco | Quality | HV | 2.08E-01 | 2.26E-01 |
| | | ε | 1.13E-01 | 1.75E+00 |
| | | IGD | 1.28E-03 | 1.36E-02 |
| | Diversity | PFS | 1.40E+03 | 3.00E+02 |
| | | S | 5.37E-01 | 1.37E+00 |
| FreeBSD | Quality | HV | 2.71E-01 | 2.71E-01 |
| | | ε | 1.34E-01 | 2.00E-01 |
| | | IGD | 7.92E-04 | 3.91E-03 |
| | Diversity | PFS | 2.87E+03 | 2.85E+02 |
| | | S | 6.40E-01 | 1.34E+00 |
| eCos | Quality | HV | 2.25E-01 | 2.83E-01 |
| | | ε | 1.08E-01 | 1.63E-01 |
| | | IGD | 1.46E-03 | 8.33E-03 |
| | Diversity | PFS | 4.15E+03 | 3.00E+02 |
| | | S | 7.23E-01 | 1.35E+00 |

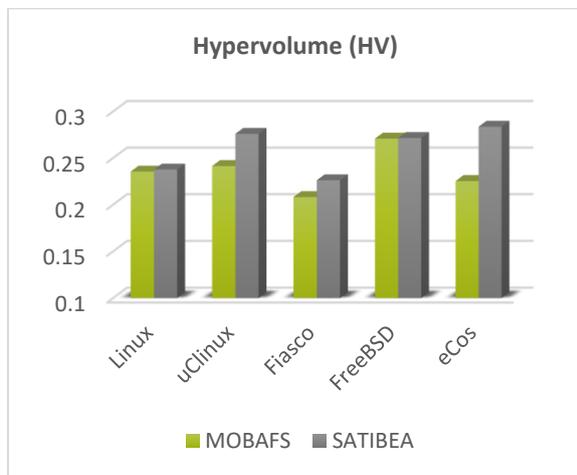

Figure 10- Hypervolume values for MOBAFS and SATIBEA

According to Epsilon metric definition in Table 1, lower value for Epsilon shows better performance. Figure 11 illustrates the Epsilon values of MOBAFS and SATIBEA and for all feature models, excluding Linux, the Epsilon values related to MOBAFS are lower than SATIBEA's Epsilon values.

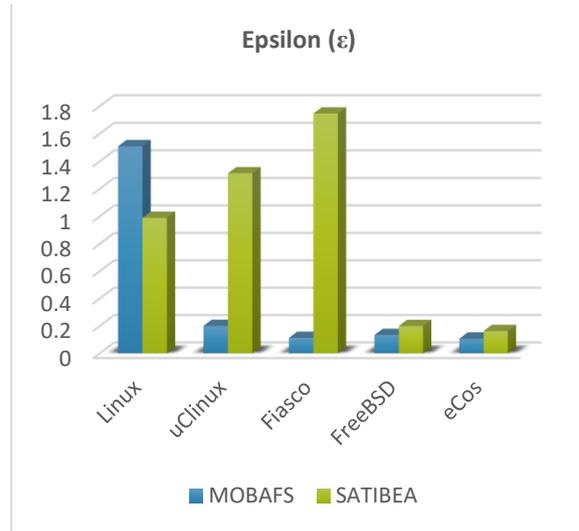

Figure 11- Epsilon values for MOBAFS and SATIBEA

Based on the definition of Inverted Generational Distance (IGD) metric in Table 1, the lower IGD, the better performance, so by considering the IGD values of MOBAFS and SATIBEA which is shown in Figure 12, MOBAFS has a marked superiority over SATIBEA in this stage due to have lower IGD values in all feature models.

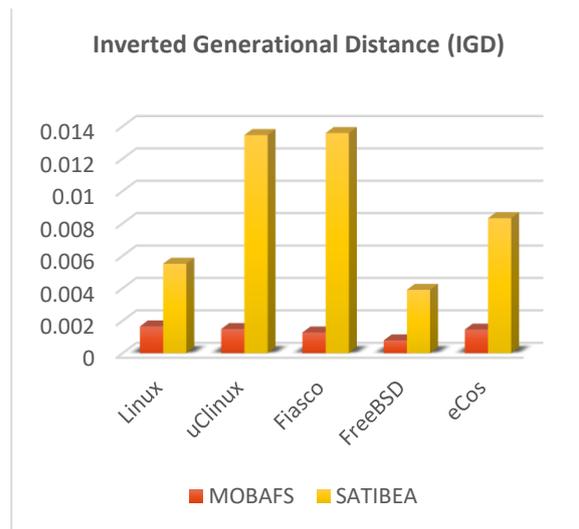

Figure 12- Inverted Generational Distance values for MOBAFS and SATIBEA

By looking at the definition of Pareto Front Size (PFS) metric, pointed in Table 1, it is clear the higher value for this metric is more favorable. By paying attention to Figure 13, MOBAFS has competitive performance in comparison with SATIBEA.

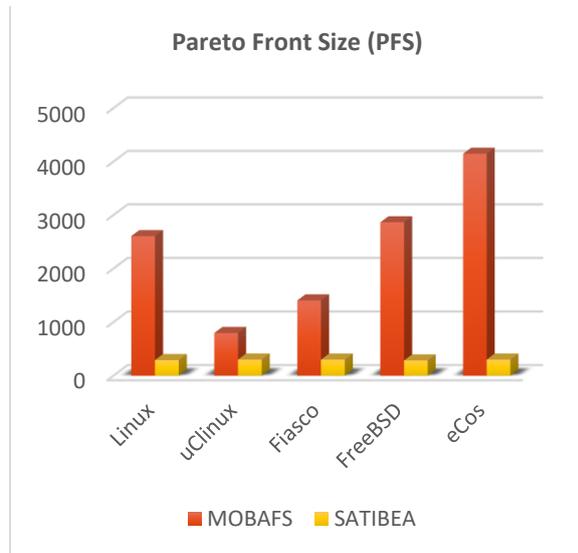

Figure 13-Pareto Front Size values for MOBAFS and SATIBEA

The Spread (S) metric, according to its definition in Table 1, shows the spread in the Pareto front, so the higher Spread indicates the more distributed solutions. Based on the comparison is illustrated in Figure 14, SATIBEA is more successful in generation of sporadic solutions.

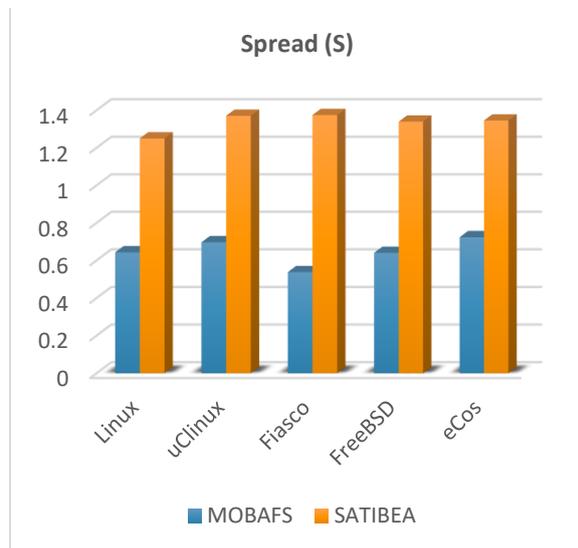

Figure 14-Spread values for MOBAFS and SATIBEA

In Figure , the relationship between Hypervolume and violated constraints in Linux feature model is illustrated. As it is obvious, with increasing in the number of violated constraints, the amount of Hypervolume will be decreased.

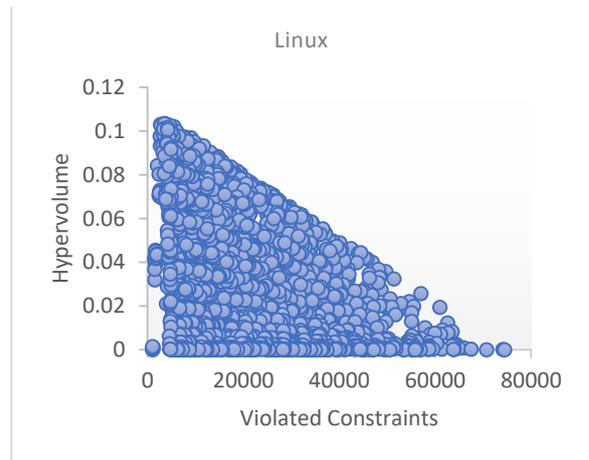

Figure 18- The relationship between Hypervolume and violated constraints

### 5.1. Statistical Test

In this work, transformed Vargha-Delaney effect size measurement [60] is applied to assess the new algorithm. As indicated by Neumann et al.[60], the mentioned non-parametric effect size test returns a $\hat{A}_{12}$ statistic which is between 0 and 1. $\hat{A}_{12} = 0.5$ shows that the two algorithms are completely equivalent; otherwise they have some difference. For instance, if $\hat{A}_{12} = 0.8$ then algorithm A overcomes algorithm B with higher values, 80% of the times.

Table 6 illustrates the $\hat{A}_{12}$ statistic to evaluate the results. Concerning HV, the most contrast is in uClinux and eCos (SATIBEA produces better results in 69.5% (1-0.305) and 77.7% (1-0.223) of the times, respectively) and for other feature models there are not significant difference between the algorithms. Regarding Epsilon, for Linux feature model, SATIBEA has better performance in 86.3% of the times. On the other hand, for other feature models, MOBAFS surpasses SATIBEA in 78.1% (1-0.219) to 100% of the times. Concerning IGD, for all feature models, MOBAFS gets better results in 87.4% (1-0.126) to 100% of the times. Regarding PFS, MOBAFS achieves more excellent results in 100% of the times, for all feature models. Concerning Spread metric, unlike PFS, SATIBEA gets highest Spread for all feature models in 100% of the times.

TABLE 6- TRANSFORMED $\hat{A}_{12}$ STATISTICAL TEST RESULTS FOR MOBAFS-SATIBEA.

| Feature Model | HV | ε | IGD | PFS | S |
|---|---|---|---|---|---|
| Linux | 0.469 | 0.863 | 0.126 | 1.000 | 0.000 |
| uClinux | 0.305 | 0.000 | 0.000 | 1.000 | 0.000 |
| Fiasco | 0.430 | 0.000 | 0.000 | 1.000 | 0.000 |
| FreeBSD | 0.492 | 0.103 | 0.117 | 1.000 | 0.000 |
| eCos | 0.223 | 0.219 | 0.000 | 1.000 | 0.000 |

### 5.2. Threats to Validity

As indicated by Lopez-Herrejon et al. [38], a common internal validity threat is adequate parameter setting. Default parameter values, which were applied by their main authors, for the two algorithms under comparison is employed. The two external threats, as mentioned in [38], are the selection of multi objective algorithms to compare and the selection of feature models. According to this point that this is for the first time that MOBAFS algorithm is used to solve a SBSE problem and SATIBEA is a strong contender in this field, so these two algorithms have been chosen. In terms of feature model, the most highly prestigious and the largest real feature models are considered which sometimes led to increase the execution time. Applying other algorithms and feature models could be a new research field.

*5.3. Discussion of Results*

In the current section, results are described in detail and classified based on the feature models. MOBAFS in 3 metrics ( i.e. Epsilon, IGD and PFS) and SATIBEA in 2 metrics (i.e. HV and Spread) have better performance. One of the main reasons for the success of MOBAFS is applying fundamental changes in MOBSO. Parallelization and Generating solutions with higher quality are examples of the innovations of this article.

## 6. CONCLUSION AND FUTURE WORK

SPLs developers prefer to generate optimal products automatically. However, selecting an optimal subset of features, with considering constraints, is an NP-hard problem. This article, for automating product derivation in SPLs, proposed a new algorithm based on MOBSO as a feature selection method. Two improvements have been done to strengthen the power of MOBAFS; Parallelization and Generating solutions with higher quality. Our new MOBAFS algorithm was compared with SATIBEA, the recent most successful algorithm, based on the largest and most prestigious feature models. Experiments showed that the new method, in most cases, is competitive with SATIBEA.

As future works, we plan to improve our new algorithm in other aspects to achieve a robust one. In addition, we would apply the new method for other SBSE problems.